\documentclass[prapplied,preprint,amsmath,amssymb,longbibliography,showpacs,showkeys,superscriptaddress,byrevtex]{revtex4-1}
\usepackage{amsmath,latexsym}
\usepackage{graphicx}
\usepackage{chngcntr}
\usepackage{amssymb}
\usepackage{amsfonts}
\usepackage{amsbsy}
\usepackage{natbib}
\usepackage{comment}
\usepackage{color}
\usepackage{float}

\usepackage[colorlinks=true, urlcolor=blue, citecolor=blue, linkcolor=blue]{hyperref}

\usepackage{silence}
\renewcommand{\vec}[1]{\mbox{\boldmath$\mathrm{#1}$}}

\renewcommand{\vec}[1]{\mbox{\boldmath$\mathrm{#1}$}}
\newcommand{\ee}{\mathrm{e}}
\newcommand{\ii}{\mathrm{i}}

\newcommand{\dblue}[1]{{\color{black} #1}}

\begin{document}

%\preprint{APS/123-QED}

\title{Active nonreciprocal cloaking for pseudo-Hermitian magnons}

\author{Dominik Schulz}
\author{Jamal Berakdar}
\affiliation{Institut f\"ur Physik, Martin-Luther Universit\"at Halle-Wittenberg, 06099 Halle/Saale, Germany}
\author{Xi-guang Wang}
\affiliation{School of Physics, Central South University, Changsha 410083, China}
\email{wangxiguang@csu.edu.cn}

\date{\today}

\begin{abstract}
	Cloaking has important applications but entails sophisticated control of signal propagation and scattering characteristics. Here, we show that invisibility for magnon signals is achievable in a non-reciprocal and electrically controlled way by engineering the magnonic channels such that they exhibit PT-symmetry. 
	This is accomplished by attaching current-carrying heavy metal contacts to the magnon waveguides and exerting fields from
	an attached bias layer. Tuning the current density in the metal layer, the magnons in this setup experience electrically controlled, compensated gain and loss due to spin-orbit torque which renders the setup PT-symmetric. The magnon dynamics is then shown to be pseudo-Hermitian with exceptional points (EPs) determined actively by an external electric field. We
 	analyze the magnon scattering from single and periodic PT-symmetric regions and identify the conditions necessary for the formation of unidirectional invisibility which can be steered by specific combinations of bias layers and current amplitudes in the heavy metal as to reach the EP. The unidirectional invisibility at EP is found to be extended for a periodic PT-symmetric region. Intrinsic damping on PT-symmetric unidirectional invisibility is shown to be marginal confirming the experimental feasibility. It is shown how the unidirectional magnons can be utilized to amplify and generate magnonic orbital angular momentum states in coupled magnetic rings demonstrating a new path for manipulating magnon propagation and processing.

\end{abstract}

\maketitle

 \section{Introduction}
 Cloaking is usually achieved by engineering and/or surrounding the scatterer by a cloak medium such that reflection of waves is diminished. Various strategies and applications of these phenomena have been discussed extensively for electromagnetic waves and photonic materials
 \cite{doi:10.1126/science.1125907,Cai2007,doi:10.1126/science.1126493,PhysRevE.72.016623,doi:10.1098/rspa.2006.1715,Chen2011,https://doi.org/10.1002/adom.201800073,https://doi.org/10.1002/adfm.202213818,Teng2024}. This work deals with invisibility of scatterer and nonreciprocity of waves in the context of magnetostatics. The waves are triggered as low-energy excitations of a magnetically ordered phase. The control and utilization of these spin waves (also called magnons) define the field of magnonics. The use of magnon signals in magnetic nanostructures has been demonstrated as a feasible route to information transmission and processing.\cite{Chumak2015,50007095,Barman2021,Pirro2021,Kruglyak2010,Serga2010,Chumak,Nikitov2015,9706176} Thereby, a prerequisite for reliable functionality is the understanding of how magnon propagate, scatter, and how to control the signals via external probes. A particularly interesting aspect of magnon motion is the nonreciprocity brought about by, for example, dipolar interactions,\cite{Chen2022,10.1063/1.4775685,PhysRevApplied.12.034012, An2013} Dzyaloshinskii-Moriya interactions,\cite{PhysRevB.106.224413,PhysRevB.88.184404,PhysRevB.93.235131,PhysRevB.102.104401} spin-orbit coupling,\cite{YU20231,PhysRevLett.132.126702} spin transfer torque,\cite{PhysRevLett.102.147202} geometric design,\cite{PhysRevB.81.140404,10.1063/1.2766842,Mruczkiewicz2013,PhysRevLett.117.227203,PhysRevB.100.104427,PhysRevB.103.144411, PhysRevApplied.18.054044}, or magnon-fluxon coupling\cite{PhysRevLett.129.117201}. The magnon nonreciprocity has been exploited in various applications such as logic devices and magnon diodes.\cite{Jamali2013, Chumak,Kruglyak2010,PhysRevApplied.12.034015, PhysRevApplied.14.034063}\\ Here, we point out a qualitatively different way to achieve field-controlled magnon non-reciprocity and complete invisibility based on parity-time (PT) symmetry. Generally in textures supporting magnons, PT symmetry is achievable by controlling the gain and loss of the magnon density as to compensate each other resulting in pseudo-Hermitian magnon propagation and the appearance of non-Hermitian degeneracies, or exceptional points (EPs).\cite{Wangxinc2020,PhysRevApplied.15.034050,PhysRevApplied.18.024073,PhysRevApplied.18.024080,https://doi.org/10.1002/aelm.202300325,Lee2015,PhysRevX.8.021066,Liu2019,PhysRevApplied.18.014003,aelm202300674}
 Here, we demonstrate the possibility of unidirectional magnon invisibility in a single magnetic waveguide through the application of an electrically tunable PT symmetric potential. The potential comprises a bias field coupled with symmetric magnon gain and loss induced by spin-orbit torque (SOT). Our analysis and simulations show that the potential functions act as a unidirectional invisible medium for magnons, enabling the control over nonreciprocal magnon reflection through electrical adjustments of the gain and loss amplitudes. The PT-symmetric unidirectional invisibility has been well demonstrated in optical systems, and extended for magnetic cavities.\cite{PhysRevLett.106.213901,PhysRevA.87.012103,PhysRevLett.123.127202,Shen14} We mainly focus on two types of PT symmetric potentials: the first comprises two adjacent regions characterized by gain and loss (cf. Fig. \ref{model1}), and the second employs a periodic arrangement of regions with alternating gain and loss (cf. Fig. \ref{model2}). These designs are achievable through attaching nano-structured metallic layers to the magnetic layer. Due to the spin Hall effect, electric currents in metal layers (hosting strong spin-orbit interaction) cause either magnon gain or loss via SOT in the adjacent magnetic layer region.\cite{Krivorotov228, Garello2013, Hoffmann2013, Collet2016} We find the necessary conditions for unidirectional invisibility and their associated control mechanisms. Intrinsic Gilbert damping does not wash out the PT-symmetry-induced unidirectional invisibility. As an application of the uncovered effect, we demonstrate with full numerical simulations, how our scheme can be used for magnon amplification and to trigger well-defined orbital angular momentum magnon states in magnetic rings. The results are potentially useful for developing new types of electrically reconfigurable nonreciprocal linear and non-linear magnonic devices. 

\section{PT-symmetric setup}
The single magnonic waveguide featuring a unidirectional invisible potential is shown in Fig. \ref{model1}. The insulating waveguides, initially magnetized along the $+x$ direction, are attached to a number of electrically separated heavy metal layers. The charge current   density $J_c$ flowing along $y$ axis in the metal layer generates the  SOT $$\vec{T} = \gamma c_J \vec{m} \times \vec{e}_x \times \vec{m} $$ with the strength $$c_{J} = \theta_{\mathrm{SH}} \frac{\hbar J_c}{2\mu_0 e t M_{\mathrm{s}}},$$ where $\vec{e}_x$ is the unit vector along $x$ axis, $M_{\mathrm{s}}$ is the saturation magnetization, $ \theta_{\mathrm{SH}} $ is the spin-Hall angle, $t$ is the thickness of magnetic layer, and $ \gamma $ is the gyromagnetic ratio. \dblue{The strength $c_J$ is linearly proportional to $J_c$. For instance, $c_J = 1648.8 $ A/m  (used below)  corresponds to $J_c = 4.4 \times 10^6 {\rm A/cm^{2}}$ in Pt.}  In cases of negative (positive) $J_c$ (or $c_J$), SOT weakens (enhances) the local effective damping of magnons in the magnetic layer which results in the formation of localized magnonic gain (loss) zones. Furthermore, attaching bias layers to the waveguide results in bias fields through interlayer exchange coupling. \\
The magnetic dynamics, and  magnons (in linear and non-linear regimes) are governed by the Landau-Lifshitz-Gilbert (LLG) equation,
\begin{equation}
 \begin{aligned}
 \displaystyle 
 \label{llg} \frac{\partial \vec{m}}{\partial t} = -\gamma \vec{m} \times \vec{H}_{\mathrm{eff}} + \alpha \vec{m} \times \frac{\partial \vec{m}}{\partial t} + \vec{T}.
 \end{aligned}
\end{equation}
 Here, $\alpha$ is the intrinsic Gilbert damping in the waveguide. The effective field $$ \vec{H}_{\mathrm{eff},p} = \frac{2 A_{\mathrm{ex}}}{\mu_0 M_{\mathrm{s}}} \nabla^2 \vec{m} + H_0 \vec{x} + H_{\mathrm{b}} \vec{x} $$ consists of the internal exchange field (with an exchange constant $ A_{\mathrm{ex}} $), the external field $ H_0 $, and the coupling field $ H_{\mathrm{b}} $ from the bias layer. For high anisotropy of the bias layer, the magnetization dynamics of this layer in the regime discussed below is irrelevant. \dblue{The bias field $H_b$ consists of contributions from the interlayer exchange coupling field and from the dipolar field. Its value is tuned  by adjusting, e.g., the spacer thickness.} While the uncovered effects are generic, for an illustration we perform numerical simulations for Yttrium Iron Garnet (YIG), meaning $M_{\mathrm{s}} = 1.4 \times 10^5$ A/m, $A_{\mathrm{ex}} = 3 \times 10^{-12}$ J/m and $H_0 = 1 \times 10^5 $ A/m. Intrinsic Gilbert damping is dependent on YIG waveguide's quality and can range from $ 10^{-5} $ to $  10^{-2} $.

 \section{Single PT-symmetric region}
\begin{figure}[htbp]
	\includegraphics[width=0.75\textwidth]{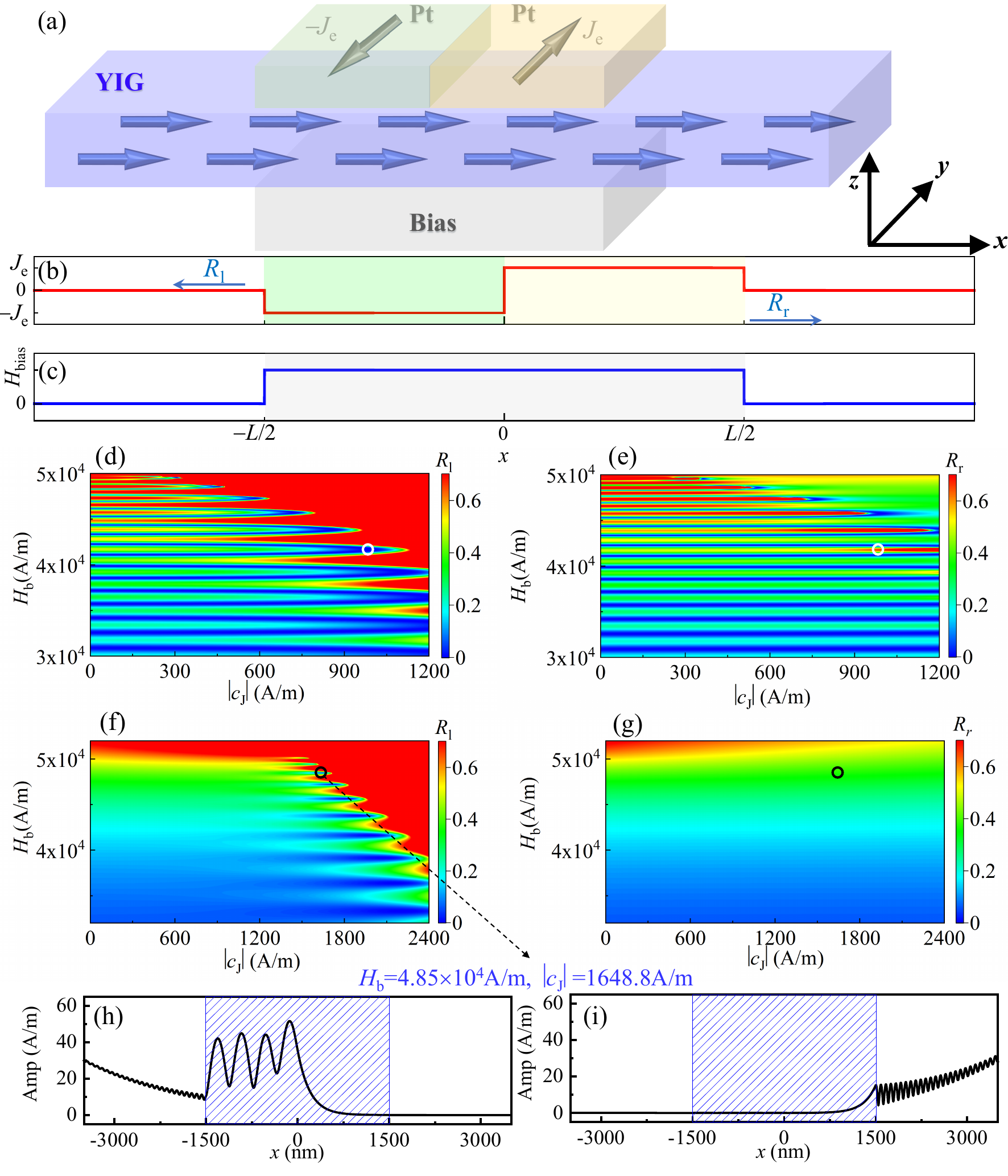}
	\caption{\label{model1} (a) Schematics  of a YIG waveguide with regions of magnonic gain and loss (blue arrows indicate remnant magnetization). Two adjacent heavy metal layers (e.g., Pt with width $L/2$) embody antiparallel charge currents $J_e$, resulting in magnon-density gain (current in  $-y$ direction) and loss ($J_e$ in $+y$ direction). The  gain/loss region is interfaced with   $L$-wide bias layer.  Spatial dependence of current (b)  and  bias field (c).  Arrow  indicates from left (right) reflected wave with reflection coefficient  $R_{\mathrm{l}} $ ($R_{\mathrm{r}} $) . (d-g) $R_{\mathrm{l}} $ and $R_{\mathrm{r}} $ dependence on $c_J\propto J_e$ and bias field $H_{\mathrm{b}}$ for $ \omega/(2 \pi) = 5.3 $ GHz with no residual damping  $\alpha = 0$ (d-e), or for $\alpha = 0.01$ (f-g). $R_{\mathrm{l(r)}}$ follows from the transfer matrix. (h-i) At $ H_{\mathrm{b}} = 4.85 \times 10^4 $ A/m and $|c_{J}| = 1648.8$ A/m (position is marked on (f)),  spatial profiles of magnons impinging from  left (h) and right (i). For $\alpha = 0.01$ amplitudes are numerically evaluated. Magnons are excited by a microwave field $h_z = h_0 \sin(\omega t)$ with $ \omega/(2 \pi) = 5.3 $ GHz and $h_0 = 1000$ A/m applied at (h) $ x = -4500$ nm and (i) $ x = 4500 $ nm.}
\end{figure}

 In Fig. \ref{model1}, the electric current's profile ( left region $J(-\frac{L}{2} \le x < 0) = -J_{\mathrm{e}}$, right region $J(0 \le x \le \frac{L}{2}) = J_{\mathrm{e}} $) yields symmetric SOT-induced gain and loss, with the bias field $H_{\mathrm{b}}$ located within the same region of ($-\frac{L}{2} \le x \le \frac{L}{2}$).\\
 To analyze full numerical simulations, we derive  analytical results in case of small transverse deviations $ \vec{m}_{\mathrm{s}} \ee^{-\ii \omega t} $, $\vec{m}_{\mathrm{s}} = (0, \delta m_y, \delta m_z)$ from the equilibrium $\vec{m}_0 = \vec{e}_x$, where $\omega$ is the magnon frequency. By defining $ \psi = \delta m_y - \ii \delta m_z $, we infer the magnon equation of motion in the form of Helmholtz equation under the linear assumption ($ |\vec{m}_{\mathrm{s}}| \ll 1 $),
 \begin{equation}
 \begin{aligned}
 \displaystyle 
 \label{sweq} \psi''(x)+\left[\frac{\omega-\omega_H-\omega_{\mathrm{b}}(x)}{\omega_k} + \ii \frac{\omega_J(x)}{\omega_k}\right]\psi(x)=0.
 \end{aligned}
 \end{equation}
 Here, we define $$\omega_k = \frac{2 (1 - \ii\alpha) \gamma A_{\mathrm{ex}}}{\mu_0 M_{\mathrm{s}} (1+\alpha^2)},$$  $$\omega_{J}(x) = \frac{(1-\ii\alpha) \gamma c_{J}(x)}{1+\alpha^2},$$   $$\omega_H = \frac{(1 - \ii\alpha) \gamma H_0}{1+\alpha^2},$$ and $$\omega_{\mathrm{b}}(x) = \frac{(1 - \ii\alpha) \gamma H_{\mathrm{b}}(x)}{1+\alpha^2}.$$ \\
 With the asymmetric current $c_J(x)$ and symmetric bias field $H_{\mathrm{b}}(x)$, the coefficient of the Helmholtz equation $$O_{\mathrm{H}} = \frac{\omega-\omega_H-\omega_{\mathrm{b}}}{\omega_k} + \ii \frac{\omega_J}{\omega_k}$$ satisfies PT symmetry condition $(O^l_{\mathrm{H}})^* = O^r_{\mathrm{H}}$ in the limit of $\alpha \rightarrow 0$, where $ O^l_{\mathrm{H}} $ and $ O^r_{\mathrm{H}} $ represent $ O_{\mathrm{H}} $ in the left and right regions respectively. 
 
 Due to the magnon scattering from the gain and loss potentials, the solution of Helmholtz equation takes the form of $$ \psi(x) = A_{\pm} \ee^{\ii k_x x} + B_{\pm} \ee^{-\ii k_x x} ,$$ where $ k_x $ is the wave number. $A_{-(+)}$ and $B_{-(+)}$ represent the forward and backward wave amplitudes at left side $z < -L/2$ (right side $ z > L/2 $). The magnon scattering is fully captured  by the transfer matrix $\hat{M}$, with $(A_+, B_+)^T = \hat{M} (A_-, B_-)^T$ (Appendix \ref{appendix:a}). From $\hat{M}$ we determine the reflection coefficients  \cite{PhysRevLett.106.213901,PhysRevLett.102.220402} from the left and right interfaces via  $$R_{\mathrm{l}} = |- \ee^{\ii k_0 L} M_{21}/M_{22}|^2,$$ and $$R_{\mathrm{r}} = | \ee^{\ii k_0 L} M_{12}/M_{22}|^2.$$  The unidirectional invisibility is defined by reflectionless scattering from left (right), meaning $M_{21} = 0 $ and $M_{12} \ne 0 $ ($M_{12} = 0 $ and $M_{21} \ne 0 $ ); these conditions we fulfill by tuning    $|c_{J}|$ and $ H_{\mathrm{b}} $.

  When Gilbert damping is neglected ($\alpha = 0$) (pseudo-Hermitian case), we can precisely determine the unidirectional invisibility condition.
As an example, setting $ H_{\mathrm{b}} = 4.17 \times 10^4 $ A/m and magnon frequency $ \omega/(2 \pi) = 5.3 $ GHz, we have the electric term amplitude $|c_{J}| = 994.8$ A/m by $M_{21} = 0 $, and here $R_{\mathrm{l}} = 0$ and $R_{\mathrm{r}} = 0.65$ represent an obvious unidirectional invisibility. To further analyze the unidirectional invisibility condition, we provided a two-dimensional distribution of $R_{l(r)}$ by varying $|c_{J}|$ and $ H_{\mathrm{b}} $ in Figs. \ref{model1}(d-e). Without SOT ($|c_{J}| = 0$), the reflections from the left and right are symmetric ($R_{\mathrm{l}} = R_{\mathrm{r}}$). $R_{\mathrm{l(r)}}$ oscillation with $H_{\mathrm{b}}$ is related to the magnon resonant transmission effect.\cite{PhysRevB.76.184419} Increasing $|c_{J}|$, every two neighboring regions with $R_{l(r)} = 0$ gradually merge into one region. The critical values of $|c_{J}|$ for merged zero reflection regions are smaller for $R_{\mathrm{l}}$. For example, at $ H_{\mathrm{b}} = 4.17 \times 10^4 $ A/m, two neighboring $R_{\mathrm{l}} = 0$ regions merge at $|c_{J}| = 994.8$ A/m (marked by the white circle in Figs. \ref{model1}(d-e)), but  still $R_{\mathrm{r}} \ne 0$. The asymmetric variation in $R_{\mathrm{l}}$ and $R_{\mathrm{r}}$ causes unidirectional invisibility. As $ R_{\mathrm{r}} = 0 $ regions merge at higher $|c_{J}|$, where $R_{\mathrm{l}} \ne 0$, unidirectional invisibility with opposite direction is generated, and here we use $ H_{\mathrm{b}} = 4.57 \times 10^4 $ A/m and $|c_{J}| = 991.6$ A/m with ($R_{\mathrm{r}} = 0$, $R_{\mathrm{l}} \ne 0$) as an example. These unidirectional invisibility conditions mark the collapse of two eigenvalues $ [1 \pm \sqrt{1-M_{11} M_{22}}]/M_{22} $ of the scattering matrix,\cite{PhysRevLett.106.213901,PhysRevLett.102.220402,MUGA2004357} indicating a spontaneous PT-symmetry breaking point, i.e., exceptional point (EP). The merged regions within Figs. \ref{model1}(d-e) also reflect this feature of EP.

In reality, the influence of finite Gilbert damping term must be included. With $\alpha \ne 0$, $k_0$, $k_1$ and $k_2$ are complex, and no numerical solution is found from $M_{21} = 0 $ or $M_{12} = 0 $, in which case we perform full-numerical simulations and observe the approximate unidirectional invisibility. See Figs. \ref{model1}(f-g), with $\alpha = 0.01$ the resonant transmission oscillations in $R_{\mathrm{l}}$ and $R_{\mathrm{r}}$ at $|c_{J}| = 0$ disappear. By comparing Figs. \ref{model1}(d) and \ref{model1}(f), at the same position of $H_{\mathrm{b}}$, it can be seen the $R_{\mathrm{l}} \approx 0$ region appears at higher $|c_{J}|$ when $\alpha = 0.01$. Note, $R_{\mathrm{r}} \ne 0$ in the whole calculation range. For example, at $ H_{\mathrm{b}} = 4.16 \times 10^4 $ A/m and $|c_{J}| = 1998.4$ A/m, approximate unidirectional invisibility ($R_{\mathrm{l}} = 0.001$, $R_{\mathrm{r}} = 0.16$) is achieved. When $H_{\mathrm{b}}$ is large enough ($H_{\mathrm{b}} > 5 \times 10^4$ A/m), $R_{\mathrm{l}}$ becomes larger and unidirectional invisibility can not be achieved. Furthermore, to validate the unidirectional invisibility under finite damping constant, we perform numerical simulations based on the LLG, as shown in Figs. \ref{model1}(h-i). The reflected magnons from the PT-symmetric region bring in fluctuations in the magnon amplitude profile. From the fluctuation amplitude at the left side (caused by $R_{\mathrm{l}}$) and right side (caused by $R_{\mathrm{r}}$), we find that the influence of left reflection is negligible compared to that from the right reflection, proving the approximate unidirectional invisibility. Here we mainly focus on the 5.3 GHz case, as different frequencies can lead to various $H_{\mathrm{b}}$ and $|c_{J}|$ under unidirectional invisibility conditions, offering extra control parameters for this mechanism.

 \section{Periodic PT-symmetric region}
 For a PT-symmetric periodic structure with $c_J(x) = -c_{A} \sin (2 k_J x)$ and $H_{\mathrm{b}}(x) = H_{\mathrm{b}A} \cos(2 k_J x)$ in the region of $ 0 \le x \le L$, the forward and backward waves inside the periodic structure are expressed in the form of $ \psi(x) = A(x) \ee^{\ii k_J x} + B(x) \ee^{-\ii k_J x} $.\cite{PhysRevLett.106.213901} Substituting this expression into Eq. (\ref{sweq}), near the Bragg point $k_x \approx k_J$, we infer the transfer matrix $\hat{M}$,
\begin{equation}
 \label{Matrixperiod} 
 \hat{M} = \begin{pmatrix}
 \cos(\beta L) + \frac{\ii \delta_k\sin(\beta L)}{\beta} &
 -\ii \frac{(\kappa + \xi) \sin(\beta L)}{\beta} \\
 \ii \frac{(\kappa - \xi) \sin(\beta L)}{\beta} & 
 \cos(\beta L) - \frac{\ii \delta_k\sin(\beta L)}{\beta}
 \end{pmatrix}.
 \end{equation}
Here, we define 
$$\kappa =\omega_{bA}/\omega_k,\, \omega_{bA} = \frac{(1-\ii\alpha) \gamma H_{\mathrm{b}A}}{1+\alpha^2},\ \xi = \omega_{JA} / \omega_k, \, \omega_{JA} = \frac{(1-\ii\alpha) \gamma c_{A}}{1+\alpha^2},     \delta_k = k_0 - k_J, $$  and \dblue{$$\beta = \sqrt{\delta_k^2 + \xi^2 - \kappa^2}.$$}

 \begin{figure}[htbp]
 	\includegraphics[width=0.75\textwidth]{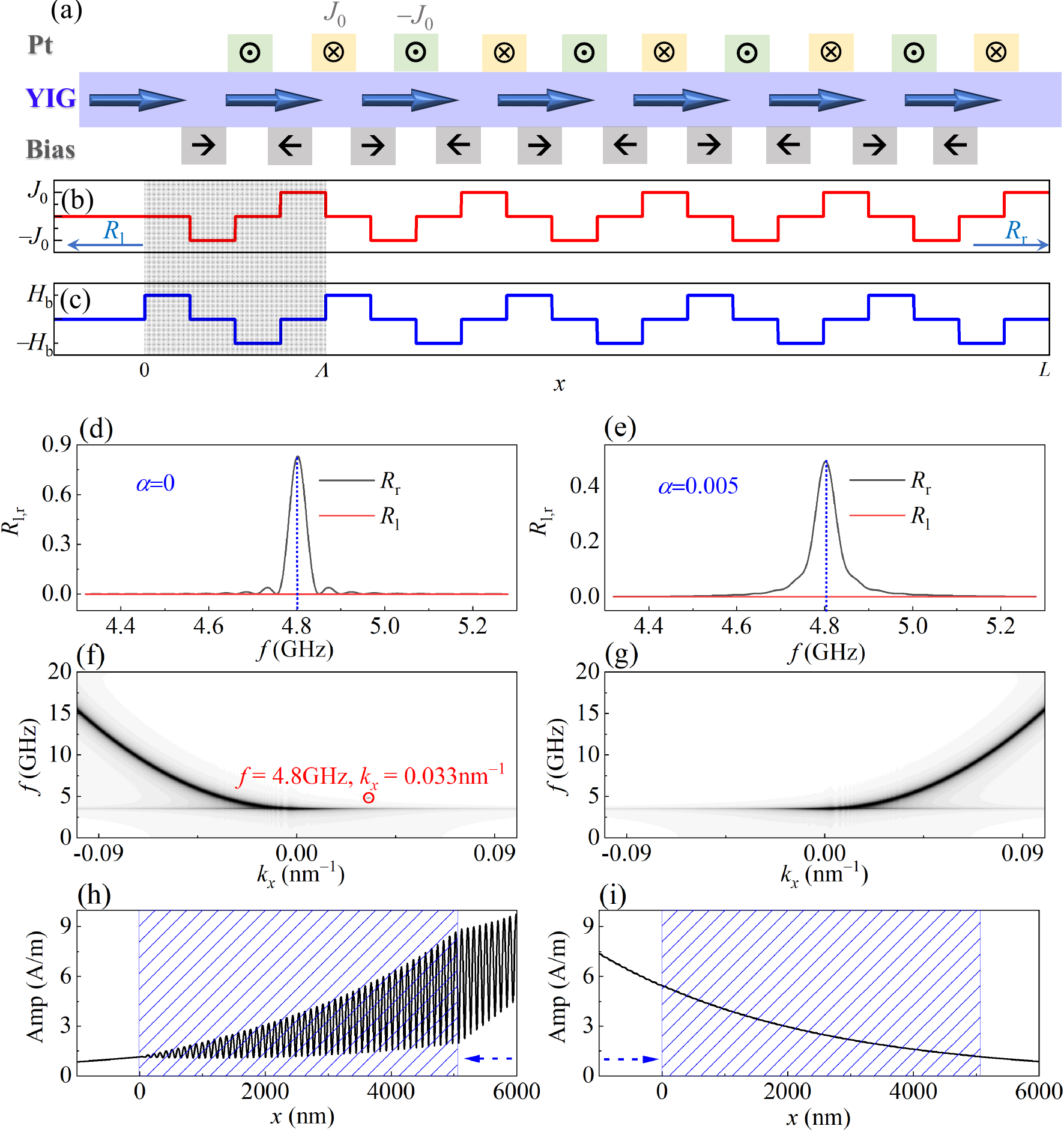}
 	\caption{\label{model2} (a) PT-symmetric  periodic magnonic structure. Neighboring heavy metal layers carry opposite currents  resulting in periodically varying gain and loss. Bias fields are periodically reversed. (b-c) Spatial distribution of the electric current density and the bias field. Gray region indicates the period $ \Lambda $, and $L$ is the periodic region length.
    (d-e) Reflection coefficients $R_{\mathrm{l}} $ and $R_{\mathrm{r}} $ as functions of magnon frequency $f$ for (d) $\alpha = 0$ and $ H_{\mathrm{b}A} = c_{A} = 400$ A/m, and (e) $\alpha = 0.005$ and $ H_{\mathrm{b}A} = c_{A} = 1000$ A/m. $R_{\mathrm{l}} $ and $R_{\mathrm{r}} $ are calculated from Eq. (\ref{Matrixperiod}) with $k_J = \pi / \Lambda $ ($ \Lambda = 96 $ nm) and $ L = 5088 $ nm. (f-g) Magnon spectra  from numerical simulations for (a)  $\alpha = 0.005$ and $ H_{\mathrm{b}A} = c_{A} = 1000$ A/m.  Magnons in frequency range 0-20 GHz are triggered by the field  $ h(t) = h_a \vec{e}_z \sin(2 \pi f_H t)/(2 \pi f_H t) $  
    right (f) or left (g)   to   the periodic structure, $h_a = 5 \times 10^3$ A/m and cutoff frequency $f_H = 20$ GHz. (h-i) At frequency $ f = 4.8 $ GHz (with wavevector $k_x = k_J$),   spatial profiles of magnon amplitudes propagating from   right (h) or left (i) sides.}
 \end{figure}

Without electric current term ($\xi = 0$), one recovers the standard reflection at a periodic Bragg structure, i.e., $R_{\mathrm{l}} = R_{\mathrm{r}} $. The periodic SOT with $\xi \ne 0$ results in an asymmetry in the $R_{\mathrm{l}}$ and $R_{\mathrm{r}}$. The asymmetry becomes most pronounced at $\kappa = \pm \xi$ causing unidirectional invisibility. For illustration, frequency ($f = \omega/2\pi$) dependent $R_{\mathrm{l}}$ and $R_{\mathrm{r}}$ are plotted in Fig. \ref{model2}(d). With $\kappa = \xi$, $R_{\mathrm{l}} $ is always 0, and $R_{\mathrm{r}} $ reaches its maximum at the Bragg point $\delta_k = 0$ for $f = 4.8$ GHz ($ k_0 = k_J = 0.033 {\ \rm nm^{-1}}$). Near the Bragg point, $R_{\mathrm{r}} $ exhibits several weaker fluctuations. Reversing the bias field or electric current ($ H_{\mathrm{b}A} = -c_{A} $), the right incident wave becomes reflectionless ($R_{\mathrm{r}} = 0$ and $R_{\mathrm{l}} \ne 0$). Different from the single PT-symmetric region above, the finite Gilbert damping $ \alpha $ doesn't affect the unidirectional invisibility condition. With $H_{\mathrm{b}A} = +(-) c_{A}$, $\kappa = +(-) \xi$ can be always satisfied independent of $ \alpha $. For example, at $\alpha = 0.005$, there still exists unidirectional invisibility at the Bragg point, as shown in Fig. \ref{model2}(e). Comparing to the case of $\alpha = 0$, the peak at this point becomes wider, and the maximum of $R_{\mathrm{r}} $ is decreased. Also, the weak fluctuations around the Bragg point are not viable anymore. Still, the unidirectional invisibility point $\kappa = \pm \xi$ in the periodic PT-symmetric region marks the collapse of two eigenvalues of the scattering matrix, representing EP of the system.
 
 We carried out numerical simulations (details in Appendix \ref{appendix:b}) for an experimentally feasible setting (Fig. \ref{model2}(a)) to confirm the above analysis. The periodic bias layers and the charge-carrying layers with period length $ \Lambda = 96 $ nm are located in the range of $ 0 \le x \le 5088 {\ \rm nm} $, and their spatial distributions are designed to approach the above periodic $c_J(x)$ and $H_{\mathrm{b}}(x)$. \dblue{To generate  periodic variation in the bias field, we suggest using the spin torque resulting from a charge-carrying metal layer attached to the bias layer,\cite{Martinez2015,Baumgartner2017} or by applying a perpendicular magnetic field pulse to control the magnetization distribution in the periodic array.\cite{PhysRevB.87.134419,Haldar2016,Wang2018}} We investigated the magnon spectra propagating from the left and right sides. See Figs. \ref{model2}(f-g), for the right-propagating ($k_x < 0$) magnon spectrum, there are reflected magnons around the Bragg point $k_x = k_J$ ($f = 4.8 $ GHz). The left-propagating magnon spectrum exists only for $k_x > 0$, indicating the existence of unidirectional reflection. We further confirm this conclusion by the spatial distribution of the magnon amplitude at 4.8 GHz in Figs. \ref{model2}(h-i). Reversing the direction of the current $c_J$ or the bias field $H_{\mathrm{b}}$ can change the direction of unidirectional invisibility (results not shown).
 
 \begin{figure}[!ht]
 	\includegraphics[width=0.9\textwidth]{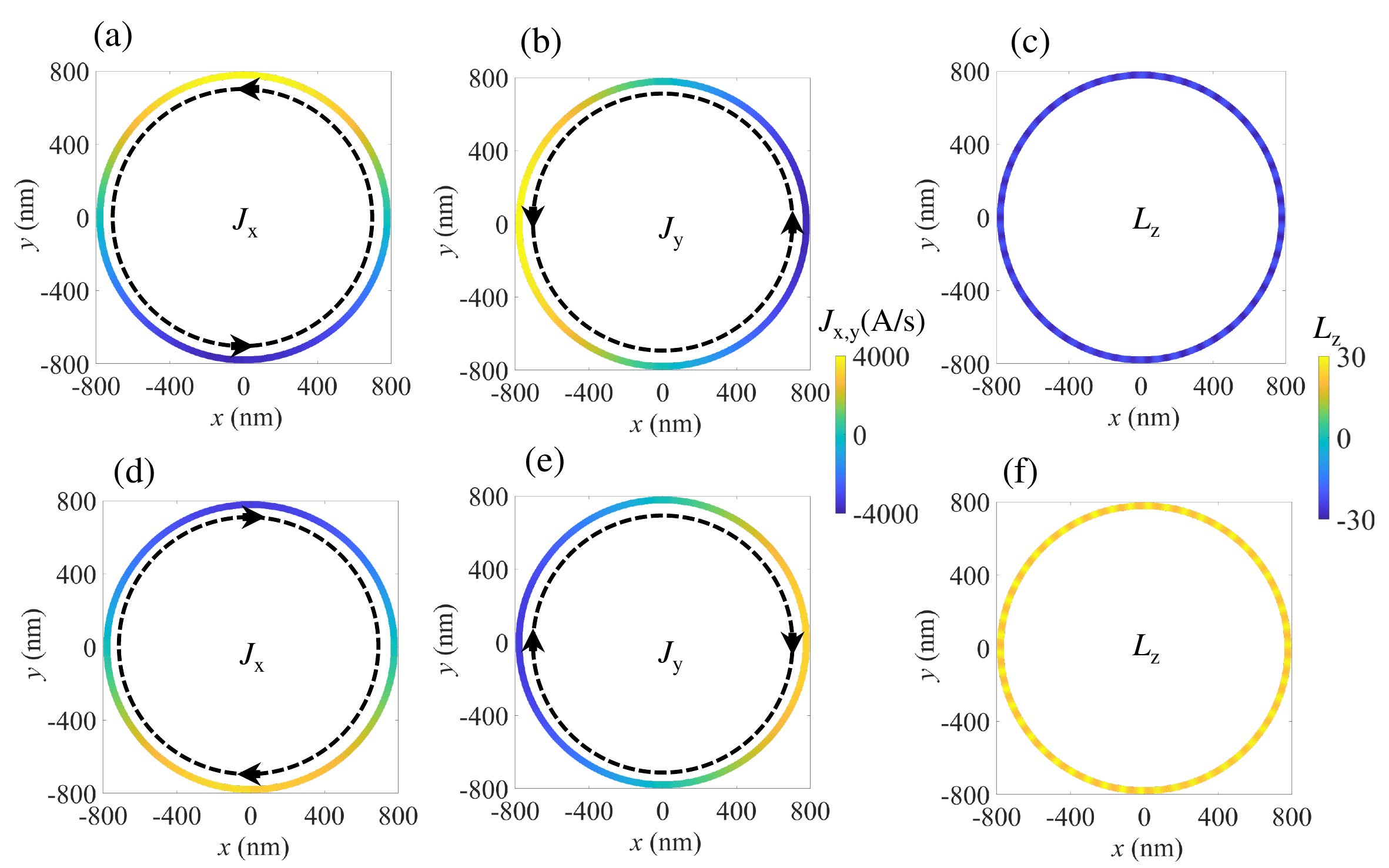}
 	\caption{\label{circle} Spatial profiles of magnonic spin currents $J_{x,y}$ and orbital angular momentum density $L_z$ (obtained from full numerical simulations) in a magnetic ring with unidirectional invisibility potential varying along the (a-c) counterclockwise and (d-f) clockwise directions. The ring is magnetized along $+z$ direction, and bias fields and SOT spin polarizations (amplitudes $ H_{\mathrm{b}A} = c_{A} = 1000$ A/m) are also applied along $z$ axis in the unidirectional invisibility potential. Magnons are excited by a microwave field $h_x = h_0 \sin(\omega t)$ with $ \omega/(2 \pi) = 4.7 $ GHz and $h_0 = 1000$ A/m located at $ (x,y) = (0, -800) $ nm.}
 \end{figure}
 
 \section{Amplification and magnon orbital angular momentum states}
As an application, we show that magnon orbital angular momentum (OAM) \cite{Jia2019} chiral state can be triggered in a magnetic ring; similar effects are expected for magnons in magnetic vortices \cite{PhysRevLett.122.097202}. The input signal can be directly applied to the ring or may stem from the coupling by proximity to a magnonic stripe. 
Such magnonic rings \cite{Wang2020,PhysRevB.108.174445,PhysRevApplied.22.014042} are feasible experimentally, and the use of chirality in magnonics \cite{10.1063/5.0068820,10.1063/5.0193495,PhysRevApplied.20.034023} is documented. 
To generate a unidirectional invisibility potential, the ring is exposed to a bias field and SOT-induced gain/loss with a period corresponding to an arc length of $\pi/25$ radians. For a ring with an average radius of 780 nm (width 40 nm), the period length is approximately 98 nm, and at the Bragg point the magnon frequency is $ \omega/(2 \pi) = 4.7 $ GHz. Simulations/discussions for magnons in rings triggered by mode-coupling to a planar-waveguide with unidirectional invisibility are in Appendix \ref{appendix:c}. Unidirectionality of  OAM magnons can be exploited to change the back-flow of magnons into the waveguide and hence to amplify the ring magnon density (results in Appendix \ref{appendix:d}).\\
Technically, introducing the unidirectional invisibility potential varying in the counterclockwise direction, the counterclockwise magnons can transmit without reflection, and those in the opposite direction are reflected near the Bragg point. The asymmetry results in more magnons in the counterclockwise direction. Such feature can be quantified by the spatial distribution of magnon spin current density \cite{Kajiwara2010} 
 $$\vec{J}_{\perp} = \frac{\gamma A_{\mathrm{ex}}}{\ii \mu_0} (\psi^* \vec{\nabla}_{\perp} \psi - \psi \vec{\nabla}_{\perp} \psi^*) = \frac{2\gamma A_{\mathrm{ex}}}{\mu_0} (m_x \vec{\nabla}_{\perp}m_y - m_y \vec{\nabla}_{\perp}m_x)$$ with $$\psi = m_x + \ii m_y.$$  In the ring magnetized along $+z$, magnons with $-z$ spin transmitting along the $+(-)x$ or $+(-)y$ directions induce the negative (positive) $J_x$ or $J_y$. As shown in Figs. \ref{circle}(a-b), the counterclockwise magnons cause positive (negative) $ J_x $ on the upper (lower) half of the ring, and on the left (right) half $J_y$ is positive (negative). Reversing the unidirectional invisibility direction can flip the magnon spin current, as proved by Figs. \ref{circle}(d-e). The unidirectionally rotating magnons result in well-defined orbital angular momentum (OAM) density  quantified by $$\vec{L} = \vec{r} \times \vec{j}$$ with $$\vec{j}_{\perp} = \frac{1}{2 \ii |\psi|^2} \left(\psi^* \vec{\nabla}_{\perp} \psi - \psi \vec{\nabla}_{\perp} \psi^*\right) = \frac{1}{m_x^2 + m_y^2} (m_x \vec{\nabla}_{\perp}m_y - m_y \vec{\nabla}_{\perp} m_x).$$ In Figs. \ref{circle}(c, f) counterclockwise and clockwise magnons correspond respectively to negative and positive $L_z$, and their averaged value are both $ |L_z| = 24.1$. Without the unidirectional potential, the counterclockwise and clockwise magnons are degenerate, meaning a zero averaged $L_z = 0$ for the whole ring. Noteworthy, our unidirectionality (and OAM) are thermally resistant (cf. Appendix \ref{appendix:e}) and can be switched on and off locally by ns current pulses in the metal contacts (switching on and off the PT-symmetry) \cite{PhysRevLett.131.186705} leading to a time varying OAM which should generate an OAM torque.
 
 \section{Conclusions}
We predicted the existence of PT-symmetry-related magnon unidirectional invisibility in engineered magnetic textures. The PT-symmetric region encompasses coupling fields from bias layers and SOTs from charge-carrying heavy metal contacts. Varying the bias field or current density allows for generation and manipulation of unidirectional invisibility, as shown analytically and numerically  for single or periodically varying PT-symmetric regions. Intrinsic magnetic damping does not lift the unidirectional invisibility. The phenomenon can be used in spatio-temporal non-linear \cite{10.1063/5.0152543}and/or chiral magnonics\cite{10.1063/5.0068820,PhysRevApplied.21.040503}, e.g., for triggering well-defined OAM states. The results demonstrate the versatility of magnonic systems for studying pseudo-Hermitian and non-Hermitian physics under experimentally feasible conditions.

 \section{Acknowledgements}
 This work was supported by the DFG through SFB TRR227, and Project Nr. 465098690, the National Natural Science Foundation of China (Grants No. 12174452, No. 12274469, and No. 12074437), and the Natural Science Foundation of Hunan Province of China (Grants No. 2022JJ20050), and the Central South University Innovation-Driven Research Programme (Grant No. 2023CXQD036).

\appendix
\counterwithin{figure}{section}

\section{Transfer matrix of the single PT-symmetric region}
\label{appendix:a}

To describe the magnon scattering from the single region with gain and loss potentials (Fig. 1 in the main paper), we write the solution of the Helmholtz equation as:  
$$\psi(x) = A_{\pm} \ee^{\ii k_x x} + B_{\pm} \ee^{-\ii k_x x},$$
where $ k_x $ is the wave number. $A_{-(+)}$ and $B_{-(+)}$ represents the forward and backward wave amplitudes at left side $z < -L/2$ (right side $ z > L/2 $). \\ 
The scattering is described by the transfer matrix $\hat{M}$ \cite{PhysRevA.87.012103},
\begin{equation}
\begin{aligned}
\displaystyle 
\label{transfer} \left[
\begin{matrix}
A_+ \\ B_+
\end{matrix}\right]=\hat{M} \left[
\begin{matrix}
A_- \\ B_-
\end{matrix}\right]
\end{aligned}.
\end{equation}
Performing the computations   we obtain the following expression for the entries of $\hat{M}$,
\begin{equation}
\begin{aligned}
\displaystyle 
\label{Matrix} M_{11} =&\, \left(-k_b^3 \ee^{-\ii k_i L} + k_c^3 \ee^{-\ii k_j L} - k_d^3 \ee^{-\ii k_k L} + k_e^3 \ee^{-\ii k_{l} L} \right)/k^3_{a}, \\
M_{12} =&\, \left[-\ii k_f^3\sin\!\left(\tfrac{k_1 L}{2}\right) \cos\!\left(\tfrac{k_2 L}{2}\right) - \ii k_g^3 \cos\!\left(\tfrac{k_1 L}{2}\right) \sin\!\left(\tfrac{k_2 L}{2}\right) - k_h^3 \sin\!\left(\tfrac{k_1 L}{2}\right) \sin\!\left(\tfrac{k_2 L}{2}\right)\right] / k^3_{a},\\
M_{21} =&\, \left[\ii k_f^3\sin\!\left(\tfrac{k_1 L}{2}\right) \cos\!\left(\tfrac{k_2 L}{2}\right) + \ii k_g^3 \cos\!\left(\tfrac{k_1 L}{2}\right) \sin\!\left(\tfrac{k_2 L}{2}\right) - k_h^3 \sin\!\left(\tfrac{k_1 L}{2}\right) \sin\!\left(\tfrac{k_2 L}{2}\right)\right] / k^3_{a},\\
M_{22} =&\, \left(-k_b^3 \ee^{\ii k_i L} + k_c^3 \ee^{\ii k_j L} - k_d^3 \ee^{\ii k_k L} + k_e^3 \ee^{\ii k_{l} L} \right)/k^3_{a},
\end{aligned}
\end{equation}
where $$ k_a^3 = k_0 k_1 k_2,\quad  k_b^3 = (k_0 + k_1)(k_0 - k_2)(k_1 - k_2)/8,$$
$$k_c^3 = (k_0 - k_1)(k_0 + k_2)(k_1 - k_2)/8,\quad  k_d^3 = (k_0 - k_1)(k_0 - k_2)(k_1 + k_2)/8,$$
$$k_e^3 = (k_0 + k_1)(k_0 + k_2)(k_1 + k_2)/8,\quad  k_f^3 = (k_0 - k_1) (k_0 + k_1) k_2/2,$$
$$k_g^3 = (k_0 - k_2) (k_0 + k_2) k_1/2,\quad  k_h^3 = (k_1 - k_2) (k_1 + k_2) k_0/2,$$
$$k_i = [2k_0 - (k_1 - k_2)]/2,\quad  k_j = [2k_0 + (k_1 - k_2)]/2,$$ 
$$ k_k = [2k_0 + (k_1 + k_2)]/2,\quad  k_{l} = [2k_0 - (k_1 + k_2)]/2,$$ where 
$$ k_0 = \sqrt{\frac{\omega-\omega_H}{\omega_k}}$$
is the wavevector outside the potential, and 
$$k_1 = \sqrt{\frac{\omega - \omega_H - \omega_{\mathrm{b}} - \ii \omega_{Ja}}{\omega_k}}$$ is the wavevector in the left region with gain, and
$$k_2 = \sqrt{\frac{\omega - \omega_H - \omega_{\mathrm{b}} + \ii \omega_{Ja}}{\omega_k}}$$ is the wavevector in the right region with loss, and $$ \omega_{Ja} = \frac{(1 - \ii \alpha) \gamma |c_{J}|}{1+\alpha^2} $$ is related to electric charge current.
\section{Technical details of the micromagnetic simulations}
\label{appendix:b}

For micromagnetic simulations using a self-developed code, we discretize the magnonic waveguide with a mesh cell size of $2 {\ \rm nm}$ and solve the LLG equation by a Dormand-Prince method (RK45) with a fixed time step of 0.05 ps. To calculate the magnon excitation, we first let the magnetization relax to a stationary state. Then, we add the magnon excitation on top of the initial stationary state. The magnon amplitude is related to the oscillation amplitude of the perpendicular component of the magnetization. We analyze the fluctuation statistics extracted from each cell magnetization during 100 ns. For calculating the magnon dispersion relations we use a two-dimensional fast Fourier transformation (FFT) $\vec{m}(\vec{k}, f) = F_2[\vec{m}(\vec{r},t)]$ on the extracted data.
\section{Magnon excitation in the magnetic ring}
\label{appendix:c}

\begin{figure}[!ht]
	\includegraphics[width=0.8\textwidth]{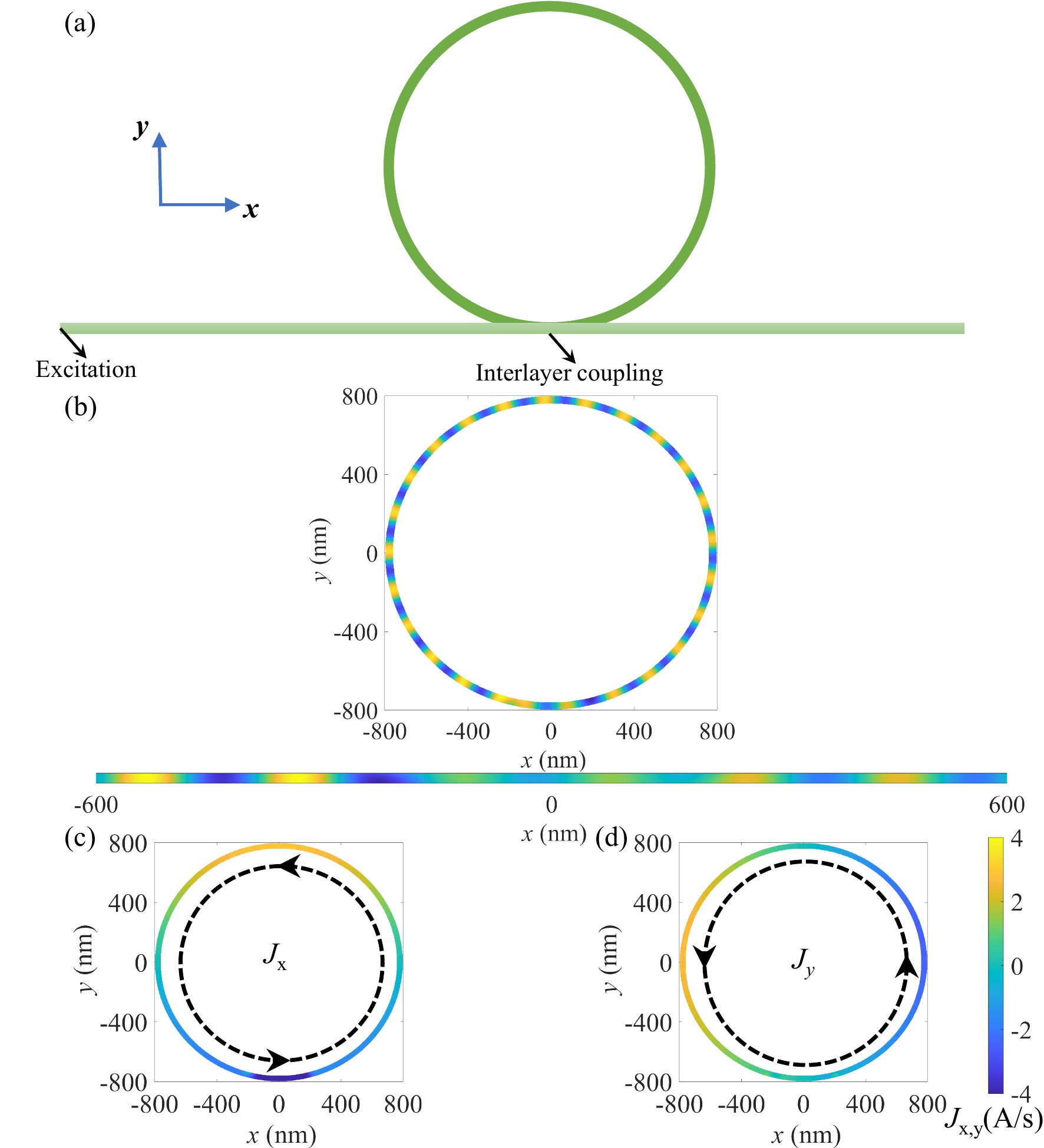}
	\caption{\label{couple} (a) Schematic for the magnon excitation in magnetic ring via mode coupling to a magnonic planar waveguide. The coupling region extends over $-250 {\ \rm nm}< x < 250 {\ \rm nm}$. (b) Distributions of magnons with frequency $f = 4.7$ GHz excited in the nanostripe and transferred to the magnetic ring. (c-d) Distributions of magnonic spin currents $J_{x,y}$ in the magnetic ring with unidirectional invisibility potential varying along the counterclockwise direction.}
\end{figure}
Magnons in magnetic ring can be excited by an external magnetic field (as in the main paper) or can be fed in via coupling to magnonic waveguides (e.g., as in Fig. \ref{couple}).  The latter case is particularly useful for integration in magnonic circuits.  We performed simulations for both cases and expect qualitatively similar results for a magnetic disc in which vortices may form.
\cite{PhysRevApplied.21.040503,10577190}
In the main text we discuss direct magnon excitations in magnetic ring. Here we focus on the case of magnetic ring proximity coupled to a nanostripe via the interlayer exchange coupling, as schematically shown in Fig. \ref{couple}.
Via mode coupling, magnons excited in the nanostripe trigger ring magnon modes. We remark that the ring-stripe mode coupling can also be achieved via dipolar interaction, as demonstrated in Ref. \cite{Wangqiham2020}, but for brevity this case is not further discussed here. For PT-symmetric dipolarly coupled spin waves we refer to Ref. \cite{PhysRevApplied.18.024080}. The magnons excited by the coupling may exhibit unidirectional transmission in a PT-symmetric potential. This feature
is evident from Fig. \ref{couple} where distributions are shown for   the magnon spin currents $J_{x,y}$ in the ring counterclockwise cloaking potential. 

\section{Magnon enhancement in the magnetic ring}
\label{appendix:d}

\begin{figure}[!ht]
	\includegraphics[width=0.9\textwidth]{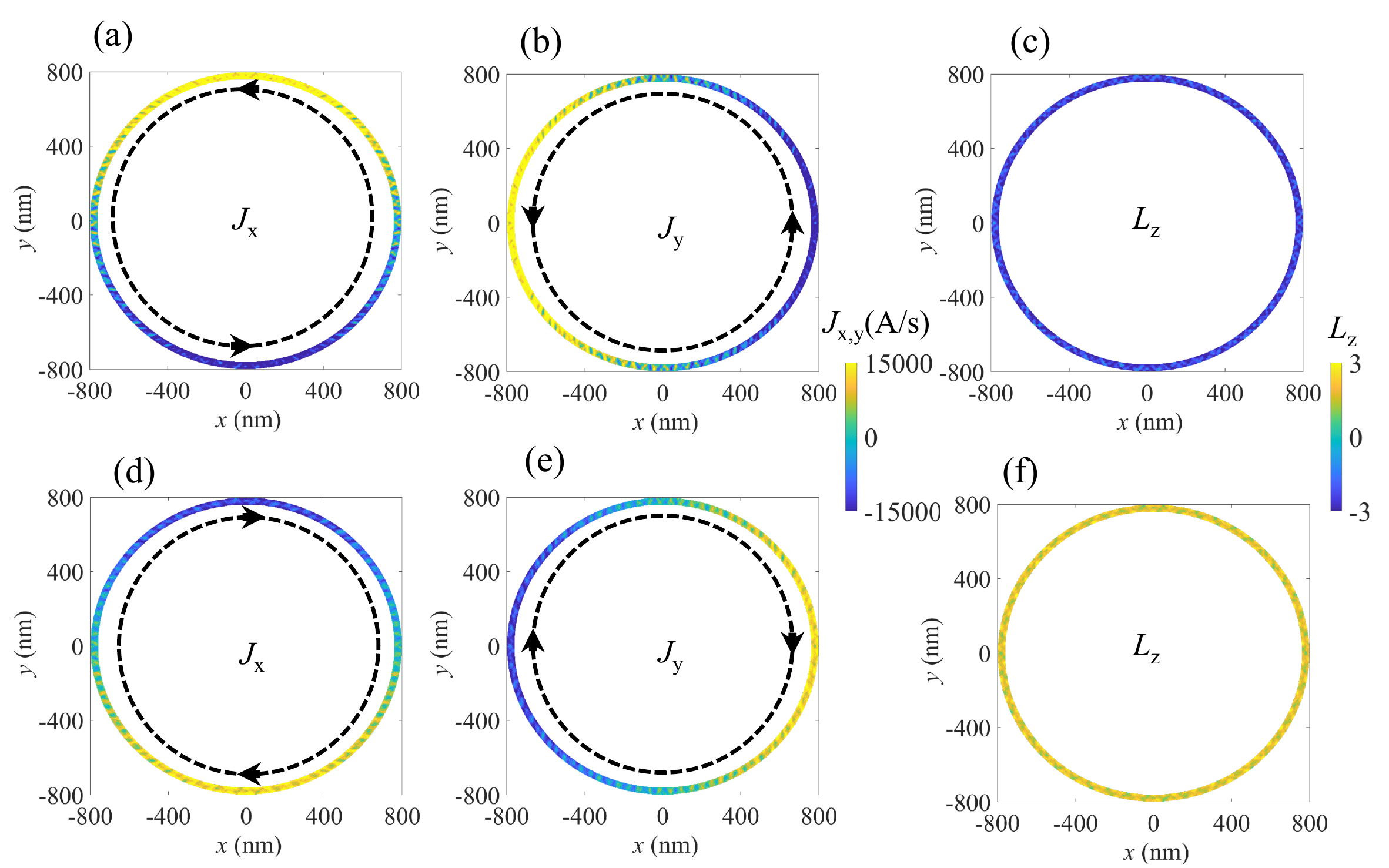}
	\caption{\label{magnondensity} Spatial profiles of magnon density $\rho = M_x^2 + M_y^2$ in magnetic ring with unidirectional invisibility potential varying along (a) counterclockwise, (b) clockwise directions, and (c) without potential. Here, magnons are excited by a microwave field $h_x = h_0 \sin(\omega t)$ with $ \omega/(2 \pi) = 4.7 $ GHz and $h_0 = 1000$ A/m located at $ (x,y) = (0, -800) $ nm.}
\end{figure}

The unidirectionality of magnonic modes can be exploited to effectively enhance the magnon density near the Bragg point
which is of interest for non-linear magnonics (captured in our case by the full numerical simulation of the magnetization dynamics). For demonstration, we
monitor the value of $ \rho = M_x^2 + M_y^2 $ to indicate changes in magnon density. As shown by Fig. \ref{magnondensity}, the magnon density is significantly enhanced in the unidirectional invisibility potential comparing with that without potential. In addition, the counterclockwise potential makes the magnon density slightly larger than that of the clockwise potential. The origin of this asymmetry is not yet well understood; one potential reason is a slight variation in magnon excitation efficiency due to changes in the magnetic field and SOT within the excitation region. A remarkable fact of using this scheme to trigger non-linear effects is that, the magnon enhancement can be controlled externally by the external probes that determine the PT symmetry, as discussed in the main text and also offers the possibility to study spatio-temporal non-linear magnonics, e.g., by using current pulses \cite{PhysRevLett.131.186705} in the metal layers and/or restricting the PT-symmetric region to parts of the magnonic channel.

\section{Thermal magnons and magnons carrying orbital angular momentum}
\label{appendix:e}

\begin{figure}[!ht]
	\includegraphics[width=0.8\textwidth]{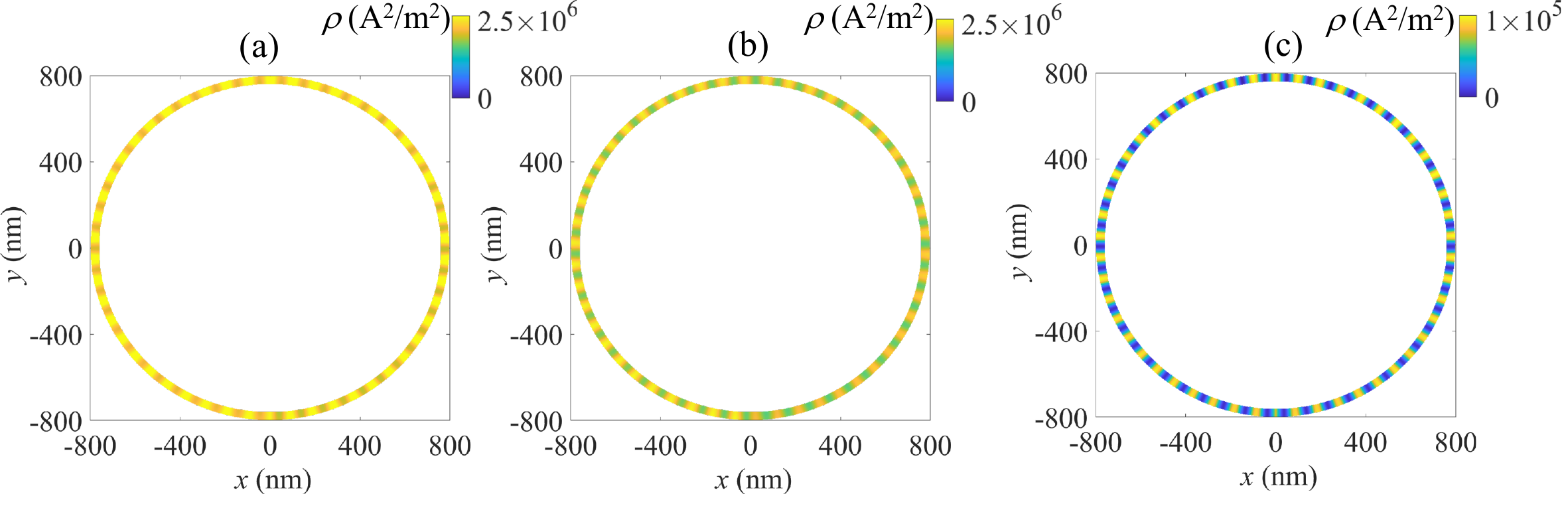}
	\caption{\label{thermal} Spatial profiles of thermal magnonic spin current $J_{x,y}$ and angular momentum density $L_z$ (obtained from full numerical simulations) in the magnetic ring with unidirectional invisibility potential varying  along the (a-c) counterclockwise and (d-f) clockwise directions. The ring is magnetized along $+z$ direction, and bias fields and SOT spin polarizations (amplitudes $ H_{\mathrm{b}A} = c_{A} = 1000$ A/m) are also applied along $z$ axis in the unidirectional invisibility potential. }
\end{figure}

In this section, we analyze the thermal magnons in the magnetic ring with unidirectional invisibility potential. In the numerical simulation based on the LLG equation, the uniform temperature $T = 30$ K is introduced via the thermal random magnetic field with the white noise correlation function $$ \langle h_i(\vec{r},t) h_j (\vec{r}', t) = \frac{2k_{\mathrm{B}} T \alpha}{\gamma M_{\mathrm{s}} V} \delta_{ij} \delta(\vec{r} - \vec{r}') \delta(t-t') \rangle .$$  Here $k_B$ is the Boltzmann constant and $V$ is the volume. Introducing the unidirectional invisibility potential varying in the counterclockwise direction, the counterclockwise magnons can transmit without reflection, and those in the opposite direction are reflected near the Bragg point. The asymmetry results in more magnons in the counterclockwise direction.  This is evident in the spatial distribution of the magnon spin current $$J_{x(y)} = \frac{2\gamma A_{\mathrm{ex}}}{\mu_0} (m_x \partial_{x(y)}m_y - m_y \partial_{x(y)}m_x).$$ In the ring magnetized along $+z$, magnons with $-z$ spin transmitting along the $+(-)x$ or $+(-)y$ directions induce the negative (positive) $J_x$ or $J_y$. As shown in Figs. \ref{thermal}(a-b), the counterclockwise magnons cause positive (negative) $ J_x $ on the upper (lower) half of the ring, and on the left (right) half $J_y$ is positive (negative). Reversing unidirectional invisibility direction can flip the magnon spin current, as proved by Figs. \ref{thermal}(d-e). Consequently, the unidirectionally revolving magnons result in a well-defined $z$ component of the (intrinsic) orbital angular momentum (OAM) density   $$\vec{L} = \vec{r} \times \vec{j}$$ with $$j_{x(y)} = (m_x \partial_{x(y)}m_y - m_y \partial_{x(y)}m_x)/(m_x^2 + m_y^2).$$ Counterclockwise and clockwise magnons correspond respectively to negative and positive $L_z$ (Figs. \ref{thermal}(c,f)).
%

%

%\section*{References}
%\bibliographystyle{naturemag}
%\bibliography{sampleBibFile}
%\newpage
%\bibliographystyle{apsrev4-1}
%\bibliography{PT-period}

\end{document}